\documentclass[aps,prx,reprint,superscriptaddress,floatfix,longbibliography]{revtex4-1}
\usepackage{graphicx,float}
\usepackage{amsmath,amssymb,mathrsfs,dcolumn}
\usepackage{gensymb}
\usepackage[usenames]{color}
\usepackage{subfigure}
\usepackage[normalem]{ulem}
\usepackage{comment}
\usepackage{hyperref}
\hypersetup{colorlinks=true, linkcolor=blue, citecolor=red, urlcolor=blue}

\begin{document}
%\title{Relaxation of squeezed magnons}
\title{Magnon bundle in a strongly dissipative magnet}
\author{H. Y. Yuan}
\affiliation{Institute for Theoretical Physics, Utrecht University, 3584CC Utrecht, The Netherlands}
\author{Jikun Xie}
\affiliation{Institute for Theoretical Physics, Utrecht University, 3584CC Utrecht, The Netherlands}
\affiliation{Shaanxi Province Key Laboratory of Quantum Information and Quantum Optoelectronic Devices, School of Physics, Xi'an Jiaotong University, Xi'an 710049, China}
\author{Rembert A. Duine}
\affiliation{Institute for Theoretical Physics, Utrecht University, 3584CC Utrecht, The Netherlands}
\affiliation{Department of Applied Physics, Eindhoven University of Technology, P.O. Box 513, 5600 MB Eindhoven, The Netherlands}
\date{\today}

\begin{abstract}
Hybrid quantum systems based on magnetic platforms have witnessed the birth and fast development of quantum spintronics. Until now, most of the studies rely on magnetic excitations in low-damping magnetic insulators, particularly yttrium iron garnet, while a large class of magnetic systems is ruled out in this interdisciplinary field. Here we propose the generation of a magnon bundle in a hybrid magnet-qubit system, where two or more magnons are emitted simultaneously. By tuning the driving frequency of qubit to match the detuning between magnon and qubit mode, one can effectively generate a magnon bundle via super-Rabi oscillations. In contrast with general wisdom, magnetic dissipation plays an enabling role in generating the magnon bundle, where the relaxation time of magnons determines the typical time delay between two successive magnons. The maximal damping that allows an antibunched magnon bundle can reach the order of 0.1, which may break the monopoly of low-dissipation magnetic insulators in quantum spintronics and enables a large class of magnetic materials for quantum manipulation. Further, our finding may provide a scalable and generic platform to study multi-magnon physics and benefit the design of magnonic networks for quantum information processing.
\end{abstract}

\maketitle

\section{Introduction}
Quantum information science utilizes the basic principles of quantum mechanics for information processing, and it has shown great potential in innovating our computing and communication technologies. Qubits, as the fundamental element to store quantum information, lie at the heart of quantum information and has been realized in superconducting systems, photonic systems, solid-state vacancies etc \cite{XiangRMP2013}. In particular, photons and phonons can also be engineered in powerful quantum circuits, even though they are continuous variable systems with an infinite number of freedoms \cite{ArrazolaNature2021}. This rapidly developing field is called continuous variable quantum information \cite{BraunRMP2005,ALexRMP2017} and has become an intriguing approach to quantum communication and quantum computation \cite{Weedbrook2012}.

Magnons, quasi-particle excitation in ordered magnets, have recently entered the territory of continuous variable quantum information. The rising field of so-called quantum magnonics manipulates the quantum states of magnons and the integration of magnon platforms with other quantum systems, including superconducting qubits, photonic cavities, nitrogen-vacancy centers, and mechanical oscillations \cite{YuanReview2022,BabakReview2022}. A hybrid magnetic system benefits from the tunability of the magnon frequency from gigahertz to terahertz regime, low relaxation rate of magnon modes, and abundant nonlinearities that even exist at room temperature. Up till now, there have been significant studies on the entanglement among magnons, photons, phonons and qubits, and various quantum states of magnons, including squeezed states \cite{ZhaoPRL2004,KamraPRL2016,LiPRA2019,KamraAPL2020,YuanPRB2021B}, single magnon states \cite{YuanPRB2020,LachScience2020}, Schr\"{o}dinger cat states \cite{SharmaPRB2021,SunPRL2021,MariosPRL2022} have been proposed. The one that has been demonstrated in the experiments is the indirect coupling of magnon and superconducting qubit mediated by the excitation of virtual photons in a three-dimensional cavity. Such coupling allows the detection of magnon excitations down to the single magnon level by reading out the state of the qubit in a delicate way \cite{LachScience2020}. However, the scalability of this hybrid platform remains a challenge. Moreover, almost all of the existing proposals use yttrium iron garnet as the magnetic medium to excite magnons because of its ultralow damping \cite{SergaJPD2010,YuanReview2022}. This rules out a large class of magnetic materials for quantum magnonics. Whether we can generate and manipulate robust magnon quantum states comparable with more diverse magnon platforms is an open question.

In this article, we study a hybrid quantum system consisting of a magnetic sphere and a superconducting flux qubit. Here the magnetic flux generated by the magnetic sphere penetrates the superconducting circuit and produces an effective coupling with both coherent and dispersive components. By driving the qubit with a frequency matching the gap of the multi-magnon energy level and qubit energy, we observe a strong multi-magnon emission, i.e., a magnon bundle, via the super-Rabi oscillations.
Interestingly, magnetic dissipation plays an enabling role in generating a sequence of magnon pairs with strong quantum correlations. The time interval between two magnons in a pair is characterized by the coherence time of the magnons, while the temporal spacing of two magnon pairs is determined by the qubit's decoherence time. The maximal damping to realize the antibunched magnon bundle can be as large as $0.1$, which is three orders of magnitude larger than that of yttrium itron garnet and thus readily enables magnetic materials ranging from insulators to metals with moderate and large damping to be useful in quantum magnonics. Moreover, our finding may provide a novel platform to study multi-magnon physics and can be extended to engineer magnonic networks for quantum information processing.

\section{Model and methodology}
\begin{figure}
  \centering
  \includegraphics[width=0.45\textwidth]{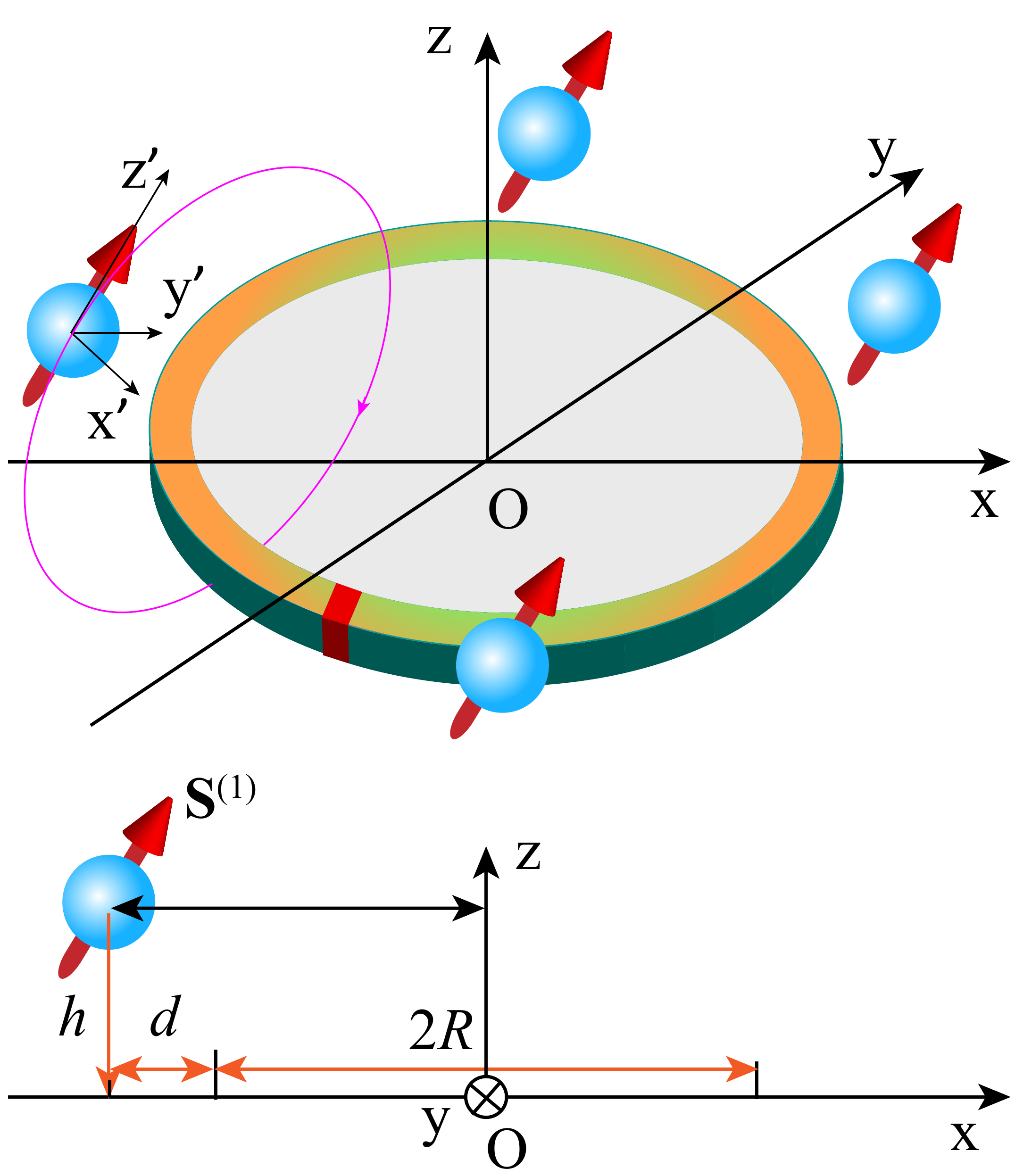}\\
  \caption{Schematic of magnetic spheres coupled to a superconducting flux qubit composing of a Josephson junction (red cuboid). Here the magnetic sphere generates a magnetic flux that penetrates through the superconducting circuit and thus couples magnon mode excited inside the magnets to the qubit state.}\label{fig1}
\end{figure}

Let us consider a hybrid quantum system composed of a superconducting flux qubit and $N_m$ magnetic spheres circulating the circuit, as shown in Fig. \ref{fig1}. The total Hamiltonian of the hybrid system is $\hat{\mathcal{H}}= \hat{\mathcal{H}}_\mathrm{m} + \hat{\mathcal{H}}_\mathrm{q} + \hat{\mathcal{H}}_\mathrm{int}$, where $\hat{\mathcal{H}}_\mathrm{m}, \hat{\mathcal{H}}_\mathrm{q}, \hat{\mathcal{H}}_\mathrm{int}$ represents the Hamiltonian for the magnet, qubit and interaction between the  magnet and qubit, respectively.
Here, the Hamiltonian of the magnetic spheres is
\begin{equation}\label{magnetc_Ham}
\hat{\mathcal{H}}_m=\sum_{l=1}^{N_m}\left (J\sum_{<ij>}\mathbf{S}_i^{(l)} \cdot \mathbf{S}_j^{(l)} - \sum_{j}\mathbf{S}_j^{(l)} \cdot \mathbf{H}^{(l)} \right ),
\end{equation}
where $\mathbf{S}_j^{(l)}$ is the $j$th spin in the $l$th magnetic sphere, and the first and second terms refer to the exchange energy between neighboring spins inside one magnet and Zeeman energy of spins subject to external field $\mathbf{H}^{(l)}$.
The flux qubit is described by the Hamiltonian \cite{Wendinarxiv2005}
\begin{equation}\label{qubit_ham}
\hat{\mathcal{H}}_\mathrm{q} = E_C \hat{N}^2 + E_J(1-\cos \hat{\varphi}) + \frac{E_L}{2} \hat{\varphi}^2,
\end{equation}
where $E_C$ is the cooper-pair charging energy, $E_J$ is the Josephson energy, and $E_L$ is the inductance of the circuit. $\hat{\varphi}$ is the phase difference of superconducting wavefunction on the two sides of the Josephson junction while $\hat{N} =-i\partial/\partial \varphi$ is the cooper pair number operator. Here the phase and charge degrees of freedom are two conjugate variables that satisfy the commutation relation $[\hat{\varphi}, \hat{N}]=i$, resembling the position and momentum operator of a particle. The interaction of the qubit with the magnetic sphere is through the magnetic flux generated by the magnets, which penetrates the magnetic circuit and modifies the potential energy of the qubit in Eq. \eqref{qubit_ham} as \cite{RusconiPRA2019}
\begin{equation}
\hat{\mathcal{H}}_\mathrm{int} = \frac{E_L}{2}\left (\hat{\varphi} -\sum_{l=1}^{N_m} \hat{\varphi}_l (\mathbf{S}^{(l)})\right)^2,
\end{equation}
where $\hat{\varphi}_l$ is the magnetic flux in the area of the circuit generated by the $l$th magnet and is defined as
$\hat{\varphi}_l=\int_{SC} \mathbf{B}^{(l)} \cdot d\mathbf{S}$ with $\mathbf{B}^{(l)}$ being the magnetic inductance generated by the $l$th magnet.

Now, we are ready to quantize the total Hamiltonian. To be clear in mathematical notation, we shall first focus on a single magnetic sphere ($N_m=1$) and then generalize it to the case of several spheres. In general, we apply a sufficiently strong magnetic field to magnetize the sphere in a uniform state. This allows us to introduce a local frame in which the axis of the magnetization is along the $z'$ axis while the $x'$ and $y'$ axes are defined in a right-hand clockwise manner, as shown in Fig. \ref{fig1}. The magnon excitation above the ground state are quantized by the standard Holstein-Primakoff (HP) transformation \cite{HPPR1940}, i.e.,
\begin{subequations}
\begin{align}
&S_i^+=\sqrt{2S-a_i^\dagger a_i}a_i, \\
&S_i^-=a_i^\dagger\sqrt{2S-a_i^\dagger a_i}, \\
&S_{z'}=S-a_i^\dagger a_i,
\end{align}
\end{subequations}
where $S_i^{\pm} = S_{x'} \pm i S_{y'}$ are the spin raising and lowering operators and we have removed the upper label ($l=1$) of spin operator for simplicity. By substituting the HP transformation into the Hamiltonian \eqref{magnetc_Ham} and transferring to momentum space via the Fourier transformation $\hat{a}_i =1/\sqrt{N} \sum \hat{a}_k e^{i\mathbf{k}\cdot \mathbf{r}}$ with $N$ being the total number of spins in the magnet, we obtain the effective Hamiltonian as $\hat{\mathcal{H}}_m=\omega_k \hat{a}_k^\dagger \hat{a}_k$. Here the magnon dispersion reads $\omega_k = H + 2ZJSk^2$ with $Z$ being hte coordinate number of the lattice.

The flux qubit is quantized by treating the double-well potential in Eq. (\ref{qubit_ham}) as two parabolas \cite{Wendinarxiv2005}, where an external flux lifts the degeneracy of the potential and generates an effective two-level system. The effective Hamiltonian reads
\begin{equation}
\hat{\mathcal{H}}_\mathrm{q} = \frac{1}{2} (\epsilon \hat{\sigma}_z + \Delta \hat{\sigma}_x),
\end{equation}
where the energy level splitting and transverse field are, respectively,
\begin{subequations}
\begin{align}
&\epsilon = 2f E_L \sqrt{\frac{6(E_J - E_L)}{E_J}}, \\
&\Delta =\sqrt{E_C(E_J - E_L)}\exp \left [ -\frac{12(E_J-E_L)^{3/2}}{E_C^{1/2} E_L} \right ].
\end{align}
\end{subequations}
Here $f = \varphi_{ex}- \pi$ is the small phase difference between external flux and $\pi$, which allows the expansion of the anharmonic potential $\cos \hat{\varphi}$ up to the quartic orders of $f$.
Due to the exponential factor in $\Delta$, it is usually much smaller than the qubit resonance frequency ($\epsilon$) and will be dropped below.

\begin{figure}
  \centering
  \includegraphics[width=0.45\textwidth]{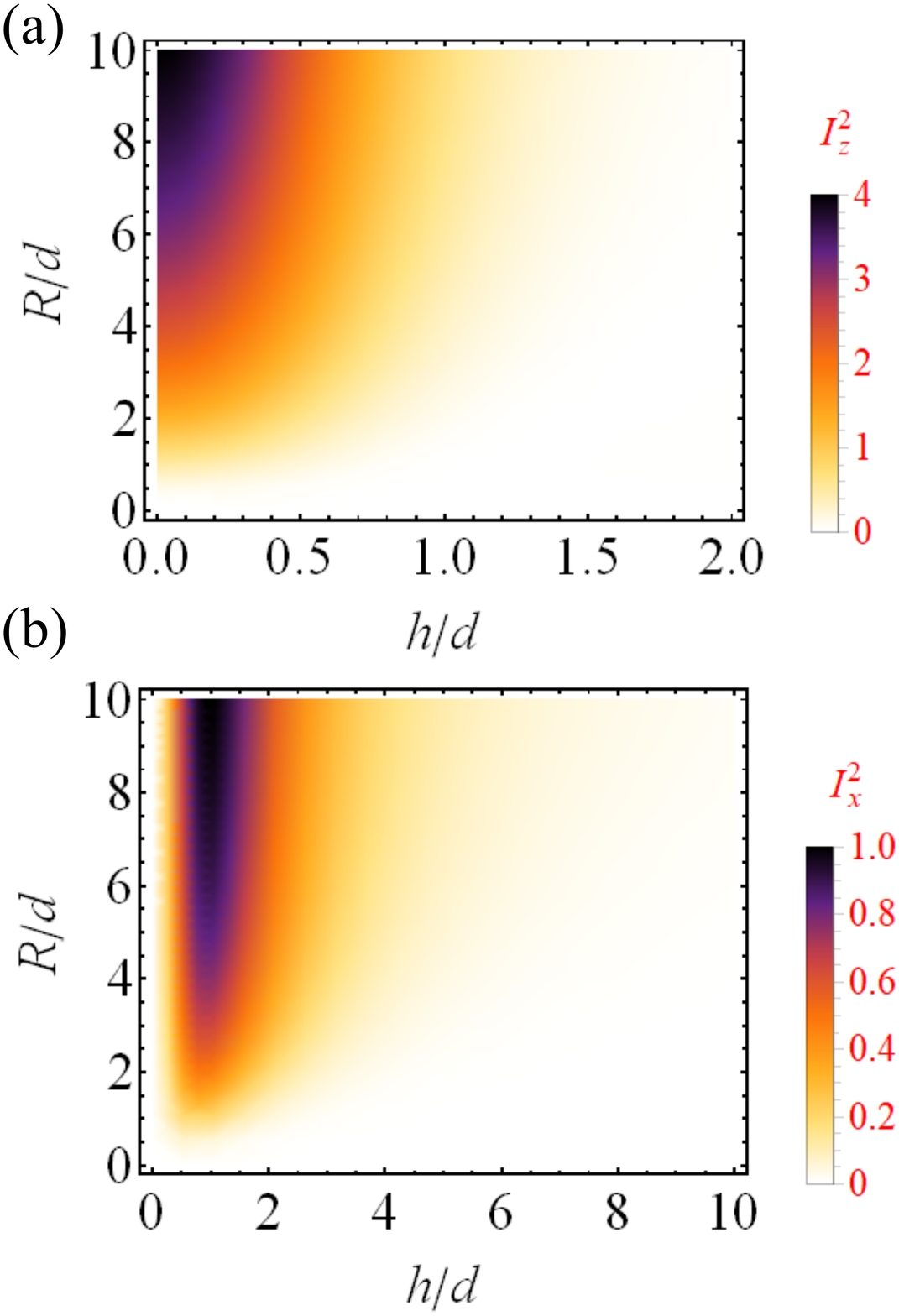}\\
  \caption{Geometric parameters of the hybrid system. $I_y =0$ for magnetic sphere locating at $(R+d,0,0)$. The coupling strength between a magnetic sphere and flux qubit is proportional to these geometric parameters as listed in Table \ref{tab1}.}\label{fig2}
\end{figure}

When the magnetic flux produced by the magnetic spheres is taken into account, it will further add an additional term to the phase difference $f$ as
\begin{equation}\label{magnetic_flux}
\varphi_m = \frac{1}{\Phi_0}\int_{SC}\mathbf{B} \cdot d\mathbf{S}= \frac{2 \pi}{\Phi_0}\frac{\hbar \gamma \mu_0}{4\pi d} \sum_i \mathbf{S}_i \cdot \mathbf{I},
\end{equation}
where $\Phi_0$ is the quantum of magnetic flux and $\mathbf{I}=(I_x, I_y, I_z)$ is a geometric vector that depends on the relative distance between the magnet and superconducting circuit, as shown in Fig. \ref{fig2}.
When the magnetic sphere is deposited as the same surface of the superconducting circuit ($h=0$), $I_x=I_y=0, I_z \approx 2$ for $R \gg d$. This implies that only the $\hat{S}_z$ component of spin is coupled to the qubit.

We note that the point dipole approximation is used in deriving Eq. \eqref{magnetic_flux}. Thus, it does not quantitatively hold when the size of the magnetic sphere $R_m$ and the distance between the magnet and superconducting circuit $d$ is comparable to the circuit size $R$.
Furthermore, for non-uniform spin-wave modes ($\mathbf{k} \neq 0$) excited in the magnet, the sum of spin fluctuations at different positions of the magnet will cancel each other. Hence, they do not contribute to the net magnetic flux. What remains is a static flux generated by the stable magnetization and the uniform fluctuation around the steady magnetization. Since the static flux can be absorbed into the term $\varphi_{ex}$,  we only have to consider the ferromagnetic resonance mode
$\mathbf{k}= 0$. This allows us to remove the subindex $i$ of spin position in the HP transformation and treat each magnetic sphere as a macrospin.

Now we write down the quantized form of the interacting Hamiltonian
\begin{equation}
\hat{\mathcal{H}}_\mathrm{int} = \hat{\sigma}_z (g \hat{a} + g^* \hat{a}^\dagger) + G\hat{\sigma}_z \hat{a}^\dagger \hat{a},
\end{equation}
where the first term and second terms denote contributions from the transverse component ($x' y'$) and longitudinal ($z'$) component of the spin fluctuations, respectively. Again, the coupling strengths $g$ and $G$, as follows from the magnetic flux (\ref{magnetic_flux}), depend on the magnetization direction and geometric factors of the hybrid system. Table \ref{tab1} shows the coupling strength as a function of geometry factors for stable magnetization along the $x$, $y$, and $z$ axes, respectively.
\begin{table}
\caption{Effective coupling strength between the flux qubit and magnetic sphere. $g_0 = 2E_L \sqrt{6(E_J-E_L)/E_J}\hbar \gamma \mu_0/(4\Phi_0d)$.}
\begin{tabular}{c|c|c}
  \hline
  \hline
  Equilibrium $\mathbf{S}_0$ & Linear $g/g_0$ & Nonlinear $G/g_0$ \\
  \hline
  $x$ & $\sqrt{2S} I_x$ & $-2I_z$ \\
  \hline
  $y$ & $\sqrt{2S} (I_z-iI_x)$ & 0 \\
  \hline
  $z$ & $i \sqrt{2S}I_z$ & $-2I_x$ \\
  \hline
\end{tabular}
\label{tab1}
\end{table}

Finally, we arrive at the effective Hamiltonian of the hybrid magnet-qubit system
\begin{equation}\label{oneMagnetHam}
\hat{\mathcal{H}} = \frac{1}{2} \epsilon \hat{\sigma}_z + \omega_m\hat{a}^\dagger \hat{a} +  \hat{\sigma}_z (g \hat{a} + g^* \hat{a}^\dagger) + G\hat{\sigma_z} \hat{a}^\dagger \hat{a}.
\end{equation}
When the qubit is driven, an additional term $\hat{\mathcal{H}}_d = \zeta( \hat{\sigma}^+ e^{-i\omega_dt} + \hat{\sigma}^- e^{i\omega_dt})$ has to be added, where $\hat{\sigma}^\pm$ are Pauli raising and lowering operators, $\zeta$ is the driving amplitude and $\omega_d$ is the driving frequency.

\section{Super-Rabi oscillation}
In this section, we shall show how the hybrid magnet-qubit system can be used to generate a multi-magnon state, or magnon bundle, by properly engineering the driving amplitude and frequency of the qubit. We first transfer to a rotation frame by performing the transformation $\hat{V}=\exp(-i\omega_d t \hat{\sigma}_z /2)$ and derive a time-independent Hamiltonian reading
\begin{equation} \label{Hybrid_ham_rotation_frame}
\hat{\mathcal{H}} = \frac{1}{2} \Delta \hat{\sigma}_z + \omega_m\hat{a}^\dagger \hat{a} +  \hat{\sigma}_z (g \hat{a} + g^* \hat{a}^\dagger) + G\hat{\sigma}_z \hat{a}^\dagger \hat{a} + \zeta \hat{\sigma}_x,
\end{equation}
where $\Delta = \epsilon - \omega_d$ is the frequency detuning between the qubit and driving microwave. To identify the role of the driving term, we perform the displacement operation $\hat{U}=\exp(-i(\eta \hat{a} - \eta^* \hat{a}^\dagger)\hat{\sigma}_z /2)$. By choosing $\eta = 2g/\omega_m$, the Hamiltonian (\ref{Hybrid_ham_rotation_frame}) is rewritten as
\begin{equation}\label{Hybrid_ham_dis_picture}
\hat{\mathcal{H}} = \frac{\Delta}{2} \hat{\sigma}_z + \omega_m\hat{a}^\dagger \hat{a}  + G\hat{\sigma}_z \hat{a}^\dagger \hat{a} + \zeta ( e^{\eta^*\hat{a}^\dagger - \eta \hat{a}} \hat{\sigma}^+ + h.c.).
\end{equation}
Here, the last term enables the magnon bundle. In general, it can be expanded to be a sum of a series of terms $g_n\hat{\sigma}^+ (\hat{a}^\dagger)^n + h.c.$ with the coupling strength $g_n=\exp(-\eta \eta^*/2) \zeta / n!$. Such coupling implies that the excitation of a multi-magnon state $(\hat{a}^\dagger)^n$ is accompanied by the qubit excitation $\hat{\sigma}^+$. To maximize this parametric excitation process, one has to match the frequency of the magnons, qubits, and drivings, i.e., $\omega_d= \epsilon - n\omega_m$ without considering the nonlinear $G$ term in the Hamiltonian. This condition allows us to tune the number of magnons. To be specific, let us study the process of $n-$magnon excitation governed by the Hamiltonian (\ref{Hybrid_ham_dis_picture}).

\begin{figure}
  \centering
  \includegraphics[width=0.45\textwidth]{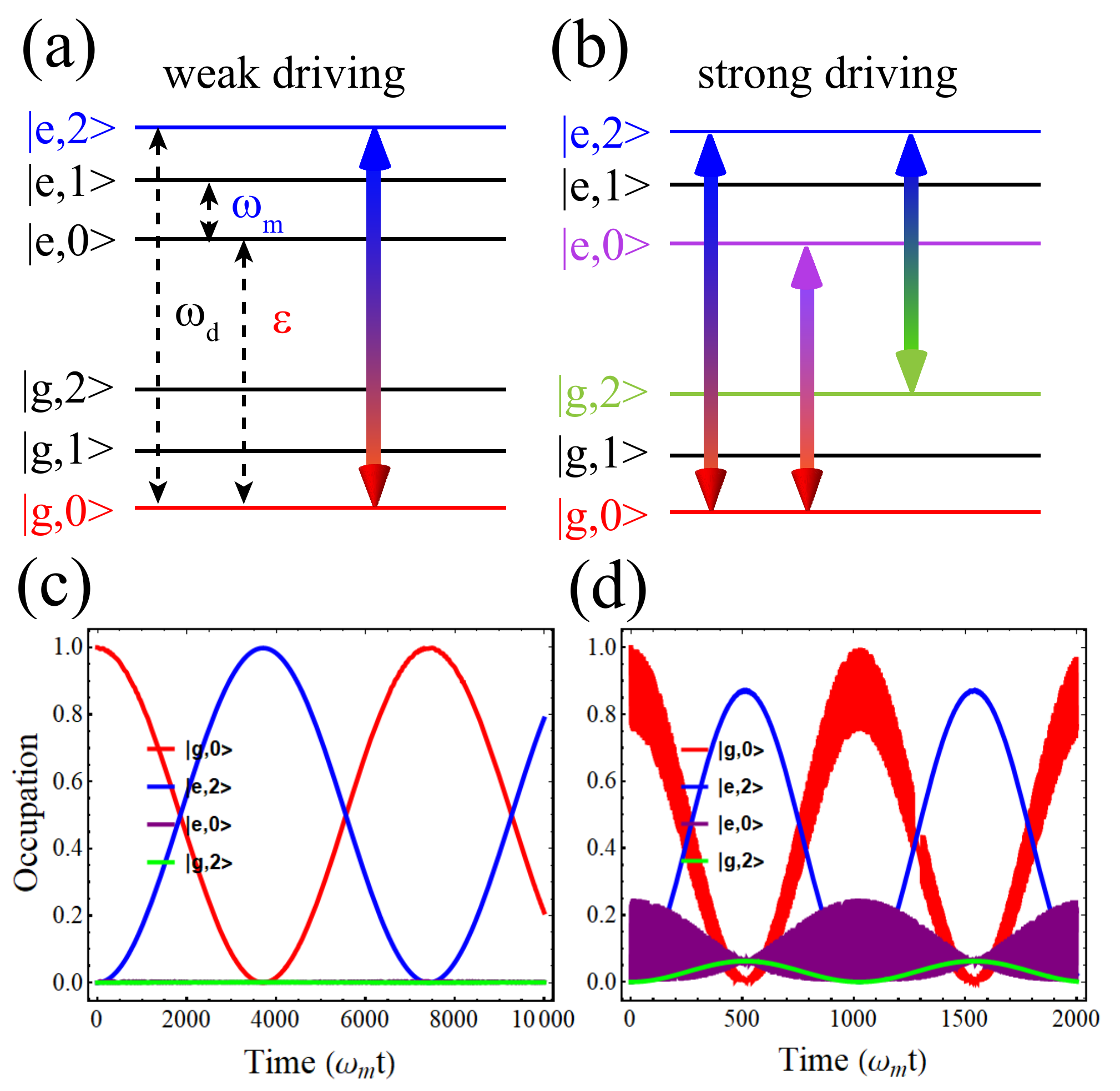}\\
  \caption{ (a-b) Schematic of energy level distribution and the dominate transitions in the weak and strong driving regimes of the hybrid magnon-qubit system. We take two magnon excitation as an example here to indicate the main transition events. (c-d) Super-Rabi oscillation in the weak and strong driving regimes. Parameters are $\zeta/2\pi = 60$ MHz for weak driving and 500 MHz for strong driving. $g=50$ MHz. $\Delta=-\sqrt{4\omega_m^2-\zeta^2}$ is taken at the two magnon resonance.}\label{fig3}
\end{figure}

Depending on the driving strength, two regimes can be distinguished as below. In the weak driving regime, the dominant transition of the system is between $| g,0 \rangle$ and $|e, n\rangle$ as shown in Fig. \ref{fig3}(a), where $|g\rangle$ and $|e\rangle$ represent the ground and excited states of the qubit and $|n\rangle$ refers to the magnon number state in Fock space. This implies that the hybrid system is in a superposition state $| \varphi \rangle = c_g|g,0\rangle + c_e|e,n\rangle$. By solving the Schr\"{o}dinger equation $i \partial_t | \varphi \rangle = \hat{\mathcal{H}}| \varphi \rangle$ under the initial conditions $c_g(0)=1, c_e(t=0) =0$, one finds
\begin{equation}
\begin{aligned}
|c_e(t)|^2 =& \frac{4 |\Omega_\mathrm{wd}|^2}{\sqrt{4 |\Omega_\mathrm{wd}|^2 + (\Delta + n\omega_m +nG)^2}}
\\ &\sin^2 \left ( \frac{t}{2}\sqrt{4 |\Omega_\mathrm{wd}|^2 + (\Delta + n\omega_m +nG)^2}\right ).
\end{aligned}
\end{equation}
At the resonance condition $\Delta + n \omega_m +nG =0$, one immediately has $|c_e(t)|^2 =\sin^2(|\Omega_\mathrm{wd}| t)$, where
the Rabi frequency reads
\begin{equation}
\Omega_\mathrm{wd} = \frac{1}{\sqrt{n!}}\zeta \exp \left [ -2\left(\frac{g}{\omega_m} \right)^2 \right ] \left(\frac{2g}{\omega_m} \right)^n,
\end{equation}
which characterizes the emission rate of the $n-$magnon state. A typical super-Rabi oscillation in the hybrid system with $n=2$ is shown in Fig. \ref{fig3}(c).

In the strong driving regime, the qubit has a larger probability to jump from the ground state to the excited state even in the absence of magnon excitations as shown in Fig. \ref{fig3}(b). Thus one has to include the energy levels $|e,0\rangle$ and $|g,n\rangle$ in the wavefunction, i.e. $| \varphi \rangle = c_g|g,0,\rangle + c_{1e}|e,n\rangle + c_{2e}|e,0\rangle + c_{3e}|g,n\rangle$. Following a similar procedure as in the weak driving case, we derive the occupation probability of $|e,n\rangle$ as
\begin{equation}
|c_{1e}(t)|^2 = \frac{\Lambda - 4 \zeta^2 - n \omega_m \Delta}{2\Omega_\mathrm{sc} \Lambda}\sin^2 (\Omega_\mathrm{sd} t),
\end{equation}
where  we assume $G=0$ to get this analytical result and the Rabi frequency $\Omega_\mathrm{sd} =\sqrt{(n\omega_m)^2 + 4 \zeta^2 + \Delta^2 + 4\Omega_\mathrm{wd}^2-2\Lambda }/2$ with $\Lambda = \sqrt{(n\omega_m)^2 (4 \zeta^2 + \Delta^2) + 16 \zeta^2\Omega_\mathrm{wd}^2}$. To maximize the maximal occupation probability, the resonance condition is now $(n\omega_m)^2 = 4 \zeta^2 + \Delta^2$, under which the Rabi frequency is rewritten as
\begin{equation}
\Omega_\mathrm{sd} = \Omega_\mathrm{wd} \frac{\Delta }{\sqrt{4\zeta^2 + \Delta^2}}.
\end{equation}
The occupation probability of $|e,n\rangle$ now becomes
\begin{equation}
|c_{1e}|^2 = \frac{1}{2} \frac{\sqrt{4\zeta^2 + \Delta^2} - \Delta }{\sqrt{4\zeta^2 + \Delta^2}}\sin^2 (\Omega_\mathrm{sd} t).
\end{equation}
Figure \ref{fig3}(d) shows the super-Rabi oscillation of the system and that the multi-magnon state are periodically emitted with a much higher rate as in the weak driving case. An alternative angle to understand this oscillation is from the dressed basis. Here a strong driving on the qubit will generate a hybrid of bare ground and excited states of qubit as $|+\rangle = -\sin \theta |g\rangle + \cos \theta |e\rangle,~|-\rangle = \cos \theta |g\rangle + \sin \theta |e\rangle$ with energy levels $\omega_{\pm}=\pm \sqrt{\Delta^2 + 4 \zeta^2}/2$, respectively. Taking into account of the magnon states, the hybrid system will oscillate between the dressed states $|-,0\rangle$ and $|+,n\rangle$ under resonant condition $\omega_+-\omega_- = n\omega_m$, which is exactly the same as the resonance condition that we have obtained by solving the Schr\"{o}dinger equation.

\section{Magnon pair generation}
In a real system, both the superconducting qubit and magnonic system will interact with the environment and are subject to relaxation and dephasing. Hence, it is meaningful to study the influence of these decoherence channels on the generation and stability of the magnon bundle.
Before discussing the influence of decoherence, let us first estimate the time scale of the hybrid system. The size of the superconducting loop is around $2R=10 ~\mathrm{\mu m}$, this gives a constraint on the size of the magnetic particle to validate the approximation as a point dipole. We assume $d=R_m= 1~\mathrm{\mu m}$, which gives an effective coupling $g/2\pi = 50~\mathrm{MHz}$, for ferromagnetic resonance frequency $\omega_m/2\pi = 1~\mathrm{GHz}$ and driving strength $\zeta/2\pi = 500~\mathrm{MHz}$. The Rabi frequency in the weak coupling regime can be readily evaluated as $\Omega/2\pi = 3~\mathrm{MHz}$ and corresponds to a period of $0.33~\mathrm{\mu s}$. This time scale is smaller than the decoherence time of a well-designed qubit, which can be at the order of several $\mathrm{\mu s}$ or even longer \cite{XiangRMP2013,BlaisRMP2021},  and a detailed discussion of the decoherence shall be done below.

\begin{figure}
  \centering
  \includegraphics[width=0.45\textwidth]{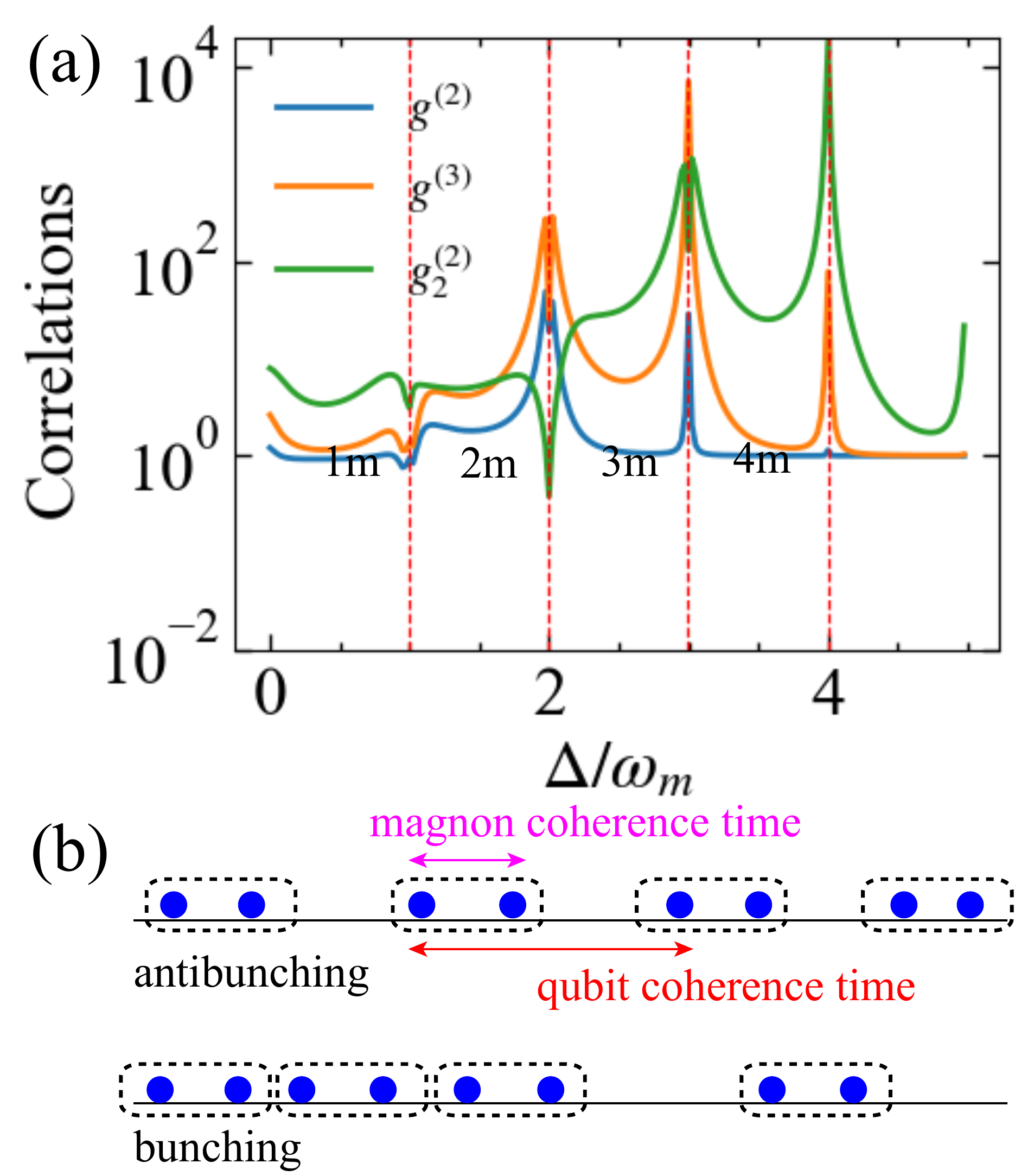}\\
  \caption{(a) Second and third order correlation functions of the steady magnon states as a function of frequency detuning between the driving and qubit modes. The dashed line indicates the position of multi-magnon process. Parameters are $ g=50~\mathrm{MHz},~\kappa_m=3\times10^{-3}\omega_m,~\kappa_q=2\times10^{-5}\epsilon,~\gamma_q=0,~\zeta=60~\mathrm{MHz}$.  (b) Illustration of magnon bundle antibunching ($g_2^{(2)}<1$) and bunching ($g_2^{(2)}>1$).}\label{fig4}
\end{figure}

To quantitatively describe the relaxation and dephasing of magnons and qubit, we consider the master equation that governs the dynamics of the system
\begin{equation}
\begin{aligned}
\frac{\partial \hat{\rho}}{\partial t}=-i[\hat{\mathcal{H}},\hat{\rho}]&+\frac{\kappa_q}{2} \mathcal{L}_{\sigma^-}[\hat{\rho}] + \frac{\gamma_q}{2} \mathcal{L}_{\sigma_z}[\hat{\rho}]\\
& + \frac{\kappa_m}{2} \mathcal{L}_a[\hat{\rho}] + \frac{\gamma_m}{2} \mathcal{L}_{a^\dagger a}[\hat{\rho}],
\end{aligned}
\end{equation}
where $\hat{\rho}$ is the density matrix of the hybrid system, and the Lindblad super-operator $\mathcal{L}$ \cite{LindbladCMP1976} is defined as $\mathcal{L}_{\hat{A}}[\hat{\rho}] \equiv 2\hat{A} \hat{\rho} \hat{A}^\dagger- \hat{A}^\dagger \hat{A} \hat{\rho}- \hat{\rho}\hat{A}^\dagger \hat{A}$. The coefficients $\kappa_q, \gamma_q,\kappa_m, \gamma_m$ characterize the relaxation and pure dephasing rates of qubit and magnon mode, respectively.
Since the flux qubit works at a temperature around 10 mK, the dephasing rate of magnons is estimated to be much smaller than the relaxation rate \cite{YuanDephase2022}. Hence, its role in the steady state may be safely neglected. To minimize the number of variables, we present results for the zero dephasing rate of the qubits ($\gamma_q=0$) below, while the general results are still valid when the qubit dephasing is included.

Let us first look at the correlation of magnons in the steady state, quantified by the $n$th-order correlation function \cite{GlauberPR1963}
\begin{equation}
g^{(n)}(\tau) = \frac{\langle \hat{a}^\dagger(0) \hat{a}^\dagger (\tau)\cdots  \hat{a}^\dagger (n\tau)  \hat{a}(n\tau) \cdots \hat{a}(\tau) \hat{a}(0) \rangle }{\langle \hat{a}^\dagger(0) \hat{a}(0) \rangle \cdots \langle \hat{a}^\dagger (n\tau)  \hat{a}(n\tau)\rangle},
\end{equation}
where $\langle \hat{A} \rangle = tr(\hat{\rho}_{ss} \hat{A})$ with $\hat{\rho}_{ss}$ being the steady density matrix of the hybrid system. Figure \ref{fig4}(a) shows the behavior of zero-delay ($\tau=0$) correlation functions $g^{(2)}$ and $g^{(3)}$ as a function of the frequency detuning between the driving and qubit mode. First, there is a strong dip in both $g^{(2)}$ and $g^{(3)}$ when $\Delta = \omega_m$, which suggests magnon antibunching. As the detuning increases around $\Delta = 2\omega_m$, a resonant dip resides inside the huge superbunching peak, indicating the existence of strong magnon correlations at the resonance. Nevertheless, $g^{(n)}(\tau)$ that characterizes the correlation of single magnons cannot give further information about the correlations of magnon pairs studied here. One may consider generalizing the second-order correlation function as \cite{MunozNP2014}
\begin{equation}
g^{(2)}_N (\tau) = \frac{\langle \hat{a}^{\dagger N}(0) \hat{a}^{\dagger N}(\tau) \hat{a}^{ N}(\tau) \hat{a}^{N}(0) \rangle }{\langle (\hat{a}^{\dagger N} \hat{a}^N) (0)  \rangle \langle (\hat{a}^{\dagger^N}  \hat{a}^N) (\tau) \rangle  },
\end{equation}
which can be reduced to the familiar form of $g^{(2)}$ for $N=1$. In general, the value of $g^{(2)}_2 (0)$ represents the statistical behavior of the magnon bundle. For $g^{(2)}_2 (0)<1$, the magnon bundle is antibunched, as shown in Fig. \ref{fig4} (b). Here a sequence of magnon pairs is evenly spaced in time. This resembles the single magnon behavior characterized by $g^{(2)}$, while the basic emission unit now becomes a two-magnon state. Indeed, the bundle correlation function $g^{(2)}_2$ shows a deep dip below one at the two magnon resonance (green line in Fig. \ref{fig4}(a)).

\begin{figure}
  \centering
  \includegraphics[width=0.49\textwidth]{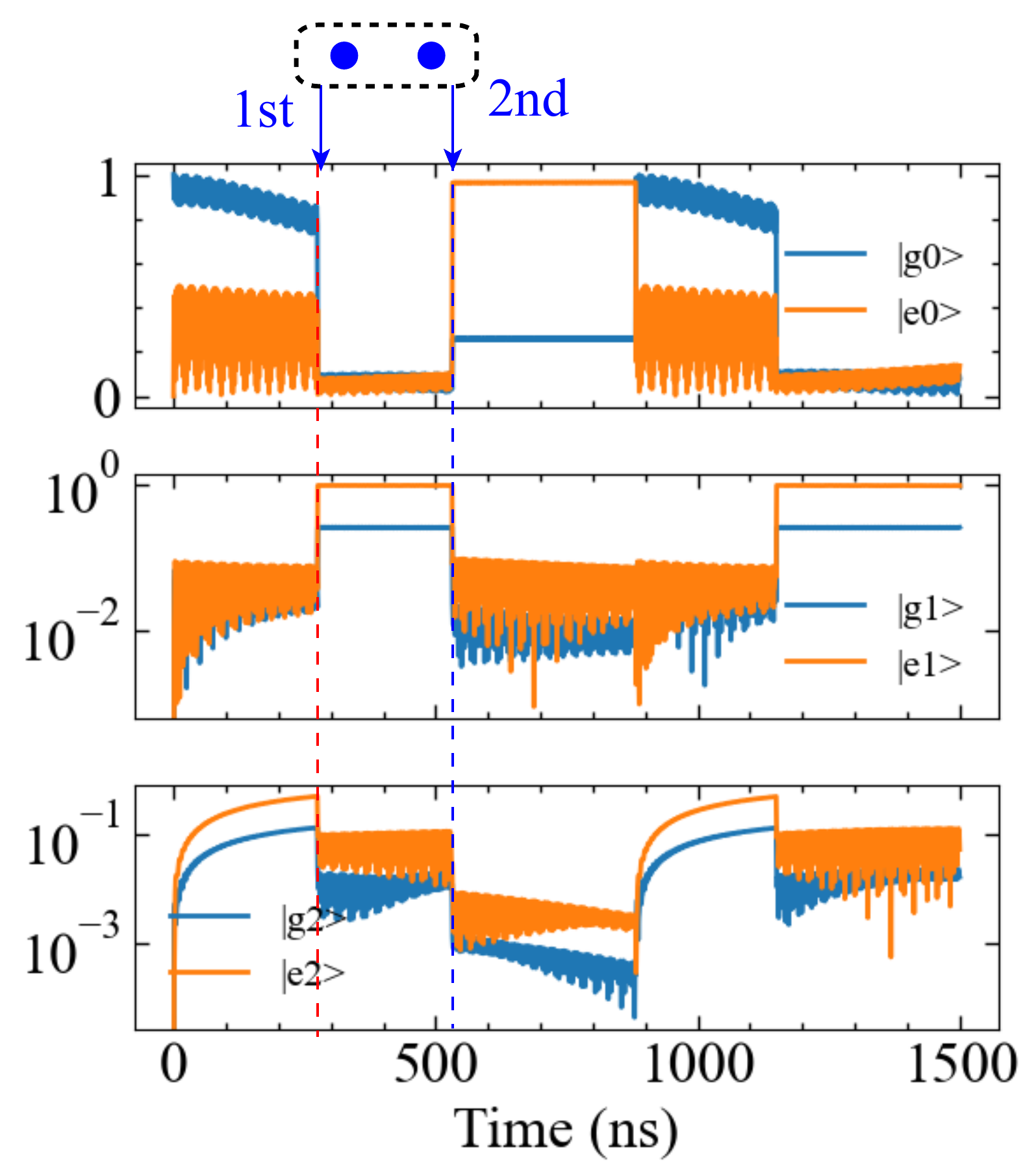}\\
  \caption{Quantum trajectory of the hybrid magnon-qubit states at the two magnon resonance. Parameters are $\kappa_m= 3\times10^{-3}\omega_m,~\kappa_q = 2\times10^{-4}\epsilon,~\zeta=500~ \mathrm{MHz}$ to increase the emission rate of two magnon bundle.}\label{fig5}
\end{figure}

To gain more insight in how the two-magnon states are emitted in the dissipative hybrid system, we employ a quantum Monte Carlo simulation, which allows us to trace the evolution of the wavefunction \cite{PlenioRMP1998,qutip2013}. In this approach, the environment is continuously monitored and generates a series of quantum jumps of the wavefunctions. Figure \ref{fig5} shows the evolution of the wavefunction at the two-magnon resonance peak, where the vertical axis can be interpreted as the probability of the system lying in the corresponding state $|g/e,n\rangle$. Initially, both the qubit and magnon are at the ground states. Before 270 ns, the states of magnon and qubit $|g/e,0\rangle$ keep oscillating. At the same time, there is a large excitation probability of the two magnon state $|e,2\rangle$, while the probability of single magnon excitation $|g/e,1\rangle$ is much smaller. At 270 ns, there appears a sudden jump of the state $|e,1\rangle$ to be close to one, which signals the emission of the first magnon. Within a time window of 250 ns, a second magnon is emitted characterized by the jump of two magnon state $|e,2\rangle$ to a significantly lower probability. Now the qubit is in the excited state ($|e,0\rangle$) and later relaxes toward the ground state at 880 ns, which also resets the high-probability of two magnon emissions (see $|e,2\rangle$ curve). Then this process repeats, and two-magnon pairs are emitted periodically.

\begin{figure}
  \centering
  \includegraphics[width=0.49\textwidth]{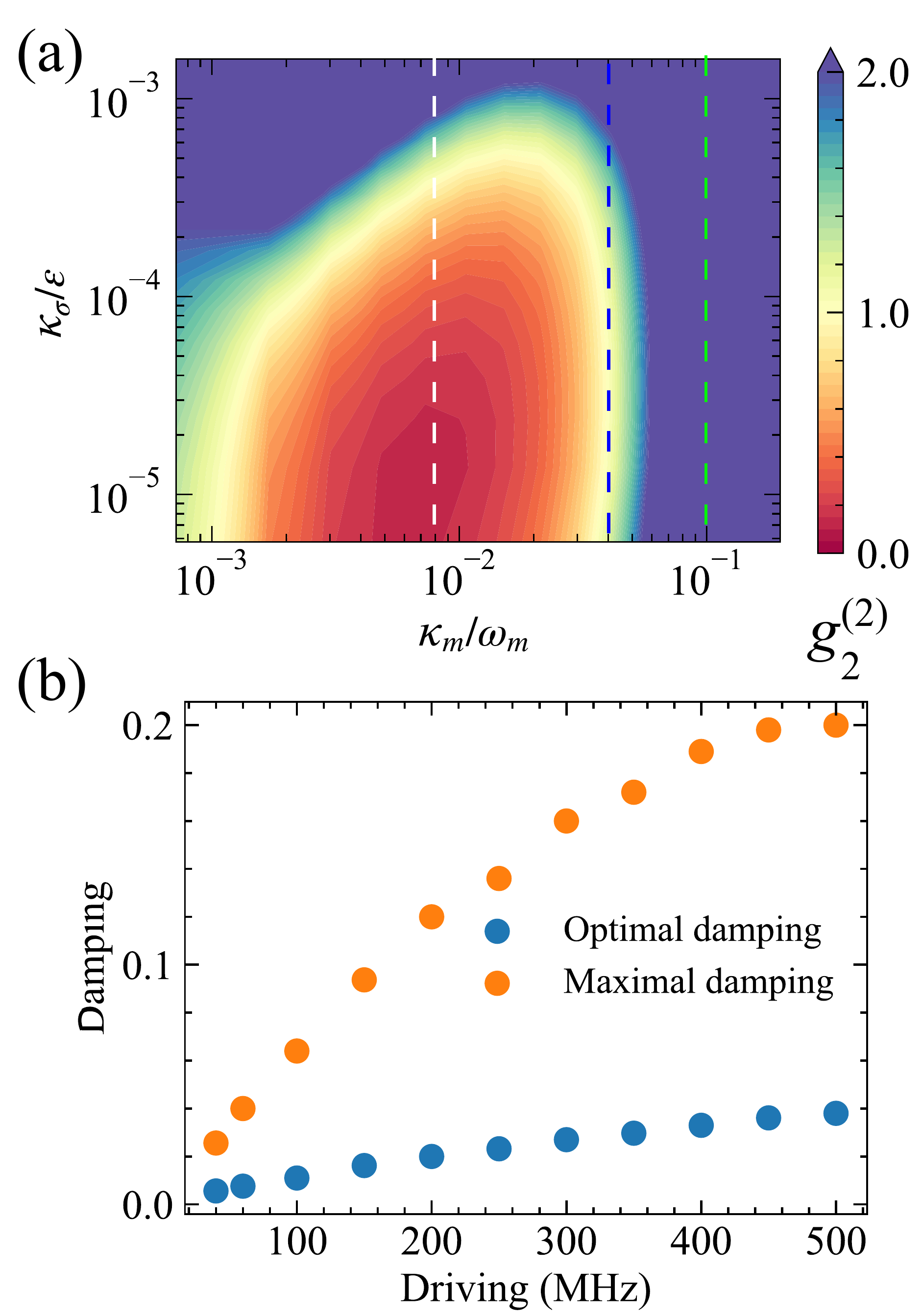}\\
  \caption{ (a) Second-order bundle correlation function $g_2^{(2)}$ as a function of magnon relaxation rate and qubit relaxation rate at the two-magnon resonance. $\Delta = - 2\sqrt{\omega_m^2-\zeta^2}, ~\gamma_q=0,~\zeta=60~\mathrm{MHz}$. (b) The optimal and maximal magnetic damping to have magnon bundle antibunching as a function of driving strength. The optimal damping is defined as the damping that the $g^{(2)}_2$ has the smallest value while the maximal damping is the position that $g^{(2)}_2$ begins to exceed one. The qubit relaxation is fixed at $\kappa_q=2\times10^{-5}\epsilon$ corresponding to the decoherence time of $10 ~\mu s$.}\label{fig6}
\end{figure}

Based on the analysis of the quantum trajectory, the magnon relaxation time characterizes the delay of the second magnon emission after the first magnon emission, while the qubit relaxation time represents the typical time to restore the hybrid system to its initial state. Figure \ref{fig6}(a) shows the phase diagram of the two-magnon emission behaviors as we tune the relaxation rate of magnons and qubits with the following features: (1) At a moderate value of magnon relaxation rate ($\gamma_m/\omega_m=0.008$, white line), the magnon bundle is antibunched when the qubit relaxation rate is below $\gamma_\sigma/\epsilon=0.0003$, which guarantees that the magnon pair is well separated in time. As the qubit relaxes faster, the system falls into the bunching regime of the magnon bundle. (2) At a considerable value of the magnon relaxation rate, the single magnon emission will play a significant role even though it is off-resonant and destroys the two magnon correlations in the steady state (green line). The maximal damping that allows for magnon bundle antibunching is located around $\gamma_m/\omega_m=0.04$, which corresponds to magnon linewidth at 40 MHz or Gilbert damping at 0.04. This maximal damping can be further increased to be of the order of 0.1 by increasing the driving of the system, as shown in Fig. \ref{fig6}(b). This enables a large class of magnetic material with moderate and large dampings to be useful for generating a magnon bundle.
%\section{Super-Rabi oscillation}

\section{Magnonic noon state}
Up till now, we have focused on the hybrid system with only one magnetic sphere. As we place two or more magnetic spheres around the superconducting circuit, more exotic quantum states can be generated. Here, we take two magnets as an example and show the generation of a magnonic NOON state. Following a similar procedure as in the one magnet case, we derive the Hamiltonian of the hybrid system as
\begin{equation} \label{multi_magnet_Ham}
\begin{aligned}
\hat{\mathcal{H}} &= \frac{1}{2} \epsilon \hat{\sigma}_z + \sum_{j=1}^2\omega_{m,j}\hat{a}^\dagger_j \hat{a}_j +  \hat{\sigma}_z \sum_{j=1}^2(g_j \hat{a}_j + g^*_j \hat{a}^\dagger_j) \\
&+\hat{\sigma}_z \sum_{j=1}^2 G_j\hat{a}^\dagger_j \hat{a}_j + \zeta(\hat{\sigma}^+ e^{-i\omega_d t} + \hat{\sigma}^- e^{i\omega_d t} ).
\end{aligned}
\end{equation}
Employing a rotation transformation $\hat{V}=\exp(-i\omega_dt \hat{\sigma}_z/2)$ and a displacement operation $\hat{U}= \exp(\hat{\sigma}_z/2 \sum_{j=1}^2(\eta_j^*\hat{a}_j^\dagger-\eta_j \hat{a}_j))$ with $\eta_j=2g_j/\omega_{m,j}$, the Hamiltonian \eqref{multi_magnet_Ham} is recast as
\begin{equation}
\begin{aligned}
\hat{\mathcal{H}} &= \frac{\Delta}{2} \hat{\sigma}_z + \sum_{j=1}^2 \omega_{m,j}\hat{a}^\dagger_j \hat{a}_j  + \sum_{j=1}^2G_j\hat{\sigma}_z \hat{a}^\dagger_j \hat{a}_j \\
 &+ \zeta( e^{  \sum_{j=1}^2 \eta_j^*\hat{a}^\dagger_j - \eta_j \hat{a}_j} \hat{\sigma}^+ + h.c.),
\end{aligned}
\end{equation}
which resembles the single magnet case Eq. (\ref{oneMagnetHam}), but now allows the excitation of a magnon bundle in each magnet. Suppose a $p-$magnon bundle and $q-$magnon bundle are respectively generated at the two magnets by properly tuning the resonance frequency, then the hybrid system may oscillate among the states $| g,0,0\rangle$, $| e,p,0\rangle$, $|e, 0,q\rangle$ and $|e, p,q\rangle$ under a weak driving. The strength of these transitions is related to the driving as $\zeta, \zeta\eta_1^p/\sqrt{p!},  \zeta\eta_2^q/\sqrt{q!},\zeta\eta_1^p\eta_2^q/\sqrt{p!q!}$. Since $\eta_j=2g_j/\omega_{m,j}$ is usually smaller than one, the joint excitation probability of the state $|e, p,q\rangle$ will be a small quantity of order $\sigma(\eta_1\eta_2)$ compared with the excitations of $| e,p,0\rangle$ and $|e, 0,q\rangle$. Therefore the wavefunction of the hybrid system may be approximated as $|\varphi\rangle = c_g| g,0,0\rangle + c_{1e}| e,p,0\rangle + c_{2e}|e, 0,q\rangle$. When the qubit is measured to be in the excited state, the wavefunction of two magnets will collapse into a NOON state $|\varphi \rangle_\mathrm{NOON}= c_{1e}| p,0\rangle + c_{2e}| 0,q\rangle$. For two identical magnets with the same number of magnon excitations ($p=q$), the coefficients satisfy the relations $c_{1e}=c_{2e}=1/\sqrt{2}$.
%\begin{figure}
%  \centering
%  \includegraphics[width=0.45\textwidth]{fig5.pdf}\\
%  \caption{}\label{fig5}
%\end{figure}
%\section{Detection}

\section{Discussions and conclusions}\label{sec_conclusion}
Our proposal launches magnonic systems as another promising platform to study multiparticle physics, in analog with their photonic and phononic counterparts \cite{MunozNP2014, BinPRL2020, LiPRR2021}. To detect the magnon excitation, one may perform state tomography on the magnon states to recover the system's density matrix and compare it with the theoretical predictions \cite{HiokiPRB2021,Xuarxiv2022}. On the other hand, one may couple the magnetic sphere to a single spin qubit, for example, a nitrogen-vacancy center, to read out the magnon states \cite{FukamiPRXQ2021}.

In conclusion, we have shown that the nonlinear interaction between a magnetic sphere and a superconducting qubit can generate a magnon bundle. When dissipative effects are considered, a sequence of magnon pairs with strong quantum correlations is generated. The time delay of two magnons inside a pair is determined by the magnon lifetime, while the two neighboring magnon pairs are well separated by the decoherence time of the qubits. This mechanism is robust over a large window of magnetic dissipation and allows for the generation of magnon quantum states in a broad class of magnetic systems with moderate and large dampings. Moreover, our methods can be generalized to manipulate magnon quantum states in a scalable magnonic network, including a periodic array of superconducting qubits and magnetic spheres.

\begin{acknowledgments}
We acknowledge helpful discussions with Marios Kounalakis. H.Y.Y. acknowledges the European Union's Horizon 2020 research and innovation programme under Marie Sk{\l}odowska-Curie Grant Agreement SPINCAT No. 101018193. J.K.X. acknowledges the support of China Scholarship Council. R.A.D. acknowledges the support as a member of member of the D-ITP consortium that is funded by the Dutch Ministry of Education, Culture and Science (OCW). R.A.D. has received funding from the European Research Council (ERC) under the European Union's Horizon 2020 research and innovation programme (Grant No. 725509).
\end{acknowledgments}

%------------------------------------------------------------------------------%
%\bibliographystyle{apsrev}
\bibliography{magnon_bundle}

%\begin{thebibliography}{99}%
%\bibitem{MetelmannPRX2015} A. Metelmann and A. A. Clerk, Nonreciprocal photon
%		transmission and amplification via reservoir engineering, \href{https://doi.org/10.1103/PhysRevX.5.021025}{Phys. Rev. X \textbf{5},021025 (2015)}.
%
%\end{thebibliography}%

\end{document}